\documentclass[pra,showpacs,showkeys, twocolumn]{revtex4}

\usepackage{amsmath}
\usepackage{amsfonts}            
\usepackage{amssymb}    

\usepackage[mathcal,mathscr]{eucal}

\usepackage{mathrsfs}

\usepackage{eufrak}

\usepackage{graphicx}
\usepackage{epstopdf}

\usepackage{dcolumn}
\usepackage{bm}

\makeatletter
\def\@dotsep{4.5}
\makeatother

\begin{document}


\title{Interaction between polar molecules subject to a far-off-resonant  optical field: \\ Entangled dipoles up- or down-holding each other}

\author{Mikhail Lemeshko}

\email{mikhail.lemeshko@gmail.com}


\author{Bretislav Friedrich}

\email{brich@fhi-berlin.mpg.de}

\affiliation{%
Fritz-Haber-Institut der Max-Planck-Gesellschaft, Faradayweg 4-6, D-14195 Berlin, Germany~}%

\date{\today}

\begin{abstract}

We show that the electric dipole-dipole interaction between a pair of polar molecules undergoes an all-out transformation when superimposed by a far-off resonant optical field. The combined interaction potential becomes tunable by variation of wavelength, polarization and intensity of the optical field and its dependence on the intermolecular separation exhibits a crossover from an inverse-power to an oscillating behavior. The ability thereby offered to control molecular interactions opens up avenues toward the creation and manipulation of novel phases of ultracold polar gases among whose characteristics is a long-range entanglement of the dipoles' mutual orientation. We devised an accurate analytic model of such optical-field-dressed dipole-dipole interaction potentials, which enables a straightforward  access to the optical-field parameters required for the design of intermolecular interactions in the laboratory.

\end{abstract}


\maketitle

\section{Introduction}

Fundamental few- and many-body physics has become a mainstay of research on ultracold gases, whose transformative effects are felt perhaps the most in condensed-matter, atomic and molecular physics~\cite{BaranovPhysRep08, LahayePfauRPP2009, TrefzgerJPB11, KreStwFrieColdMolecules}. 
Of particular interest are {\it dipolar} ultracold gases consisting either of paramagnetic atoms or of polar molecules~\cite{KreStwFrieColdMolecules}, as both are amenable to manipulation by external magnetic or electric fields. Compared with paramagnetic atoms, polar molecules offer not only electric dipole moments -- and hence a dipole-dipole interaction that exceeds the magnetic one by several orders of magnitude~\cite{LahayePfauRPP2009} -- but also vibrational and rotational structure. 

Examples of the widened scope of studies in few- and many-body physics that ensued from the use of ultracold polar molecules include:  engineering new 2D quantum phases by exploiting microwave transitions between molecular rotational states~\cite{BuchlerZollerPRL07, MicheliZollerPRA07}; construction of Hubbard-like Hamiltonians with independently controllable two-body and tree-body interaction terms~\cite{BuchlerNatPhys07}; and  simulation of any permutation-symmetric lattice spin models with open-shell molecules~\cite{MicheliNatPhys06}.

It was shown in the early 1980s that an intense far-off-resonant laser field can induce a highly-controllable retarded induced dipole-dipole interaction between atoms or molecules, which decays as $1/r$, $1/r^2$, or $1/r^3$, depending on the interatomic or intermolecular separation $r$ and  on the wavevector of the optical field~\cite{CraigThiruBook, ThiruMolPhys80}. The retarded interaction generates peculiar effects in atomic  Bose condensates, such as ``gravitational self-binding,'' rotons, as well as density modulations leading to a supersolid-like behavior~\cite{ODellPRL00, GiovanazziPRL02, ODellPRL03}. 

Recently, we presented a promising method for manipulating the interaction potentials between a pair of polar molecules with far-off-resonant light~\cite{LemeshkoOpticalShort11}. The method is based on the triple-combination of the electric dipole-dipole, anisotropic polarizability, and the retarded induced dipole-dipole interactions and offers a wide tuneability range of the intermolecular potentials that it generates. Herein, we provide a detailed account of how the combined interaction comes about, identify its short- and long-range behavior, and characterize the orientational entanglement of the dipole pair: although there is no net orientation of the pair, the dipoles are instantaneously oriented parallel or antiparallel with respect to one another depending on the state of the composite system, i.e., they ``up- or down-hold each other.''  We augmented our analysis by devising an accurate model of the triple-combined interaction which allows to readily access the optical-field parameters required for achieving preordained intermolecular interactions in the laboratory. 
 
In Sec.~\ref{Sec:OneMolecule} we describe the interaction of a single polar and polarizable molecule with a far-off-resonant optical/laser field, and in Sec.~\ref{Sec:IntermolInt} the interaction between a pair of polar molecules subject to a far-off resonant optical field. In Sec.~\ref{Sec:EffPots} we demonstrate that an optical field gives rise to new types of intermolecular potentials, which exhibit a crossover from an inverse-power decay at short intermolecular separations to an oscillating long-range behavior, and whose parameters can be  varied by tuning the intensity and frequency of the laser field.  We show  that for a wide range of field intensities and molecular parameters, the problem can be described by an exactly solvable two-level model, which leads to simple analytic expressions for the effective potential energy surfaces. We exemplify the treatment by evaluating optically-induced interactions between pairs of either $^{85}$Rb$^{133}$Cs or $^{40}$K$^{87}$Rb  molecules, widely employed in experiments with ultracold polar gases~\cite{NiJinYeNature2010, SageDeMillePRL05}. The main conclusions of this work are summarized in Sec.~\ref{Sec:Conclusions}.

\section{A molecule in a far-off-resonant laser field}
\label{Sec:OneMolecule}

We consider a diatomic molecule with a dipole moment $d$ and polarizability components, $\alpha_\parallel$ and $\alpha_\perp$, parallel and perpendicular to the molecular axis.  In a far-off-resonant radiative field of intensity $I$, the molecular rotational levels undergo a dynamic Stark shift, given by the Hamiltonian~\cite{CraigThiruBook}
\begin{equation}
	\label{Hi}
	H = B \mathbf{J}^2 - \frac{I}{2 c \varepsilon_0} e_j e_l^\ast \alpha^{\text{lab}}_{jl}(k),
\end{equation}
with $B$ the rotational constant, and $\alpha^\text{lab}_{jl}(k)$ the dynamic polarizability tensor of the molecule in the laboratory frame. The term ``far-off-resonant'' implies that the frequency of the optical field is far removed from any molecular resonance, and is much larger than the inverse of both the rotational period and the pulse duration (if a pulsed laser is used). 
We note that at the far-off-resonant wavelengths usually employed in alignment and trapping experiments ($\sim$1000 nm), the dynamic polarizability $\alpha_{ij} (k)$ approaches its static limit, $\alpha_{ij} (0)$, for a number of molecules, e.g.\ CO, N$_2$, and OCS. However, this is not the case for alkali dimers, such as KRb and RbCs, which possess low-lying excited $^1\Sigma$ and $^1\Pi$ states. Virtual transitions to these states contribute to the ground-state dynamic polarizability, rendering it a few times larger than the static value~\cite{KotochigovaDeMille10, DeiglmayrDulieuJCP08}.

\begin{figure}
\includegraphics[width=8.8cm]{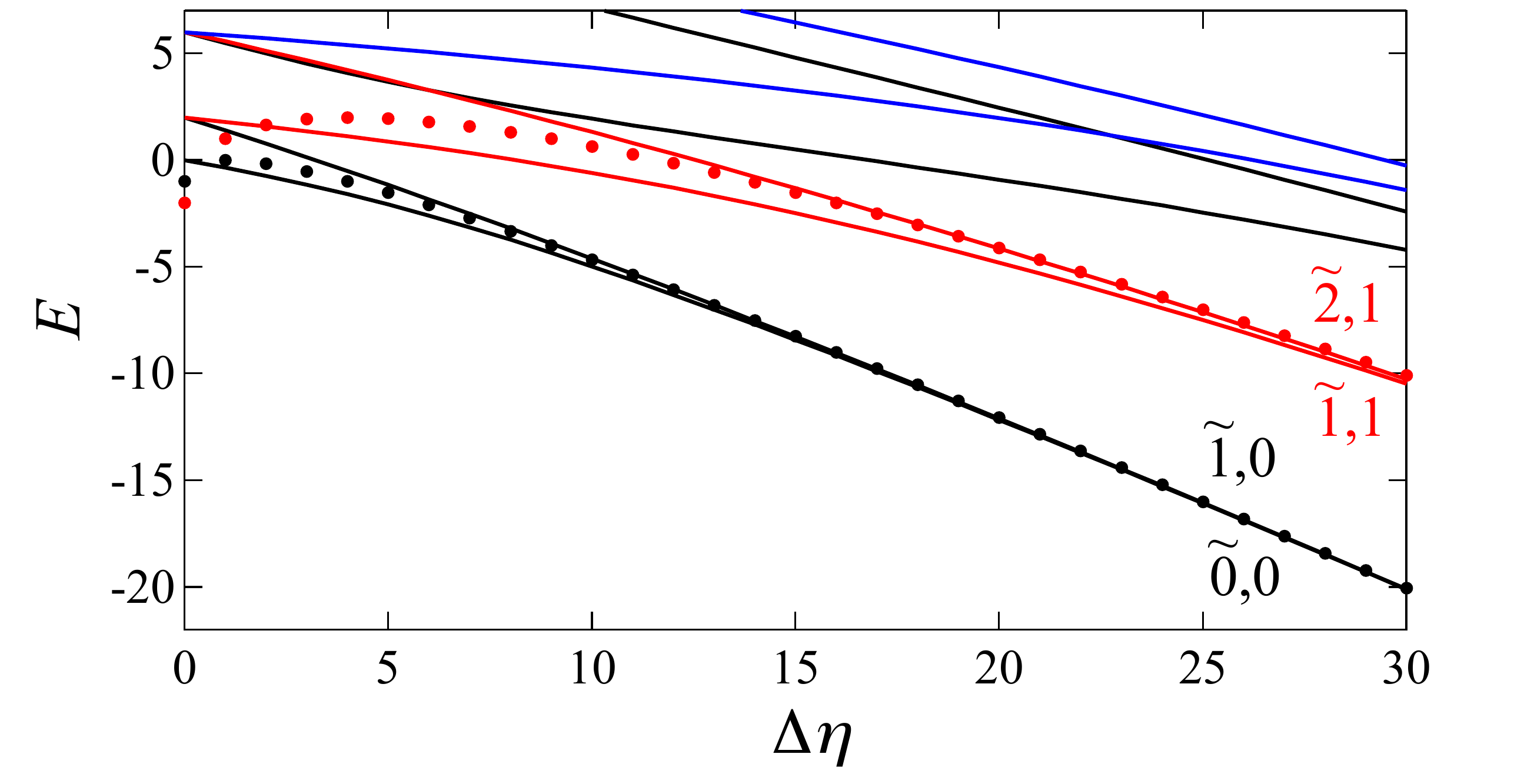}
\caption{\label{fig:AC_Stark} Dependence of the lowest energy levels of a molecule in a far-off-resonant laser field on the field-strength parameter $\Delta \eta$, with energy expressed in units of the rotational constant $B$. Different colors correspond to $M=0$ (black), 1 (red), and 2 (blue). Two lowest tunneling doublets are labeled as $\tilde{J}, M$ and the dots show their energies in the strong-field approximation. }
\end{figure}

 We assume a laser beam propagating along the positive $Y$ direction with the wavevector $\mathbf{k} = k \mathbf{\hat{Y}}$, and linear polarization, $\mathbf{\hat{e}}$ along the $Z$ axis, $\mathbf{\hat{e}}=\mathbf{\hat{Z}}$. Given that the only nonzero polarizability components in the molecular frame are $\alpha_{zz}=\alpha_\parallel$ and $\alpha_{xx}=\alpha_{yy}=\alpha_\perp$, and using $B$ as a unit of energy, Hamiltonian~(\ref{Hi}) can be recast as:
\begin{equation}
	\label{Halpha}
	H =  \mathbf{J}^2   - \Delta \eta (k) \cos^2 \theta - \eta_\perp (k),
\end{equation}
where $\theta$ is the polar angle between the molecular axis and the polarization vector of the laser field. The dimensionless interaction parameter $\Delta \eta(k)$ is defined by
\begin{equation}
\label{deltaeta}
	\Delta \eta(k) \equiv \eta_\parallel(k) - \eta_\perp (k)
\end{equation}
with  
\begin{equation}
	\label{DeltaEta2}
	\eta_{\parallel,\perp} (k) \equiv  \frac{\alpha_{\parallel, \perp} (k) I}{2\varepsilon_0 c B}.
\end{equation}
We note that eq.~(\ref{Halpha}) was derived in Refs.~\cite{FriHerPRL95, FriHerJPC95} using the semiclassical approach and the rotating wave approximation.  All rotational levels exhibit a constant shift of $\eta_\perp$,  given by the second term of eq.~(\ref{Halpha}), which will be omitted hereafter.

The polarization vector of an optical field defines an axis of cylindrical symmetry, $Z$. The projection, $M$, of the angular momentum $\mathbf{J}$ on $Z$ is then a good quantum number, while $J$ is not. However, one can use the value of $J$ of the field-free rotational state, $Y_{J, M} (\theta, \phi)$, that adiabatically correlates with the hybrid state as a label, denoted by $\tilde{J}$, so that $|\tilde{J}, M; \Delta \eta \rangle \to Y_{J, M}$ for $\Delta \eta \to 0$. For emphasis, we also label the actual values of $\tilde{J}$ by tilde, so that, e.g., $\tilde{0}$ stands for $\tilde{J}=0$.
The induced-dipole interaction, eq. (\ref{Halpha}), preserves parity, hybridizing states with even {\it or} odd $J$'s,
\begin{equation}
	\label{PendularState}
	|\tilde{J}, M; \Delta \eta \rangle = \sum_{J} c_{J M}^{\tilde{J}, M} (\Delta \eta ) Y_{J M} , \hspace{0.2cm} J+\tilde{J} \hspace{0.15cm} \text{even},
\end{equation}
and therefore results in aligning the molecules in the laboratory frame. Aligned molecules possess no space-fixed dipole moment, in contrast to species oriented by an electrostatic field.

Figure~\ref{fig:AC_Stark} illustrates how the dynamic Stark effect affects the rotational levels of a diatomic molecule and the angular `shape' of the resulting pendular states, eq.~(\ref{PendularState}). A far-off-resonant optical field of sufficiently large intensity leads to formation of ``tunneling doublets" -- closely lying states of opposite parity with same $M$ and $\vert \Delta \tilde{J} \vert  = 1$~\cite{FriHerPRL95}. Due to their proximity, the opposite-parity doublet states can be efficiently mixed by extremely weak electrostatic fields, lending them strong but opposite orientation in the laboratory frame~\cite{FriHerJCP99, FriHerJPCA99, HaerteltFriedrichJCP08}.

In the strong-field limit, $\Delta \eta \to \infty$, the eigenenergies of Hamiltonian~(\ref{Halpha}) are given by~\cite{FriHerJPCA99, FriHerZPhys96, FriHerJPC95}:

\begin{widetext}
\begin{align}
	\label{StrongFieldE1}
	      E= 2 (\Delta \eta)^{\frac{1}{2}} - \Delta \eta + 2\tilde{J}  \Delta \eta^{\frac{1}{2}} + \frac{M^2}{2} - \frac{\tilde{J}^2}{2} - \tilde{J} - 1,& \hspace{0.3cm} \text{for $\tilde{J} - |M|$ even,} \\
	\label{StrongFieldE2}
	      - \Delta \eta + 2\tilde{J}  \Delta \eta^{\frac{1}{2}}+\frac{M^2}{2} - \frac{\tilde{J}^2}{2} - \frac{1}{2},&  \hspace{0.3cm} \text{for $\tilde{J} - |M|$ odd},
\end{align}
\end{widetext}
from which it follows that the energy gap between adjacent tunneling doublets increases proportionately to $2\Delta\eta^{1/2}$, while the tunneling doublet splitting  decreases as
\begin{equation}
 \Delta E = |E| \exp [a - b\Delta \eta^{1/2}],
 \end{equation}  
with $a=3.6636$ and $b=2$ for the lowest $\tilde{0},0-\tilde{1},0$ doublet~\cite{FriHerJPCA99}.  The strong-field limit, also included in Fig.~\ref{fig:AC_Stark} and shown by dots, becomes a good approximation of the eigenenergies for the lowest doublet at  $\Delta \eta \gtrsim 15$. As we demonstrate below, at sufficiently large $\Delta\eta$, the interaction between two ground-state molecules can be described within the lowest tunneling doublet.

\section{Intermolecular interactions}
\label{Sec:IntermolInt}

\subsection{Dipole-dipole interaction}

In the absence of external fields, two polar molecules 1 and 2 interact via the dipole-dipole interaction,
\begin{equation}
	\label{DDpot}
	V_\text{dd} (r) = \frac{\hat{\mathbf{d}}^{(1)}_j \hat{\mathbf{d}}^{(2)}_l}{r^3} (\delta_{jl} - 3 \hat{\mathbf{r}}_j \hat{\mathbf{r}}_l),
\end{equation}
where $\mathbf{\hat{d}}^{(1,2)}=\mathbf{d}^{(1,2)}/d$ are unit dipole moment vectors of the molecules, $d\equiv |\mathbf{d}^{(1)}|=|\mathbf{d}^{(2)}|$, $\hat{\mathbf{r}}$ is the unit vector which defines the direction of the intermolecular axis, and 
\begin{equation}
	\label{r0}
	r_0 = \left( \frac{d^2}{4 \pi \varepsilon_0 B} \right)^{1/3}
\end{equation}
is introduced as a unit of length. In the field-free case, the interaction between two ground-state polar molecules has an isotropic asymptotic behavior, $V_\text{dd}(r) = - 1/(6 r^6)$.

Eq.~(\ref{DDpot}) can be recast in terms of spherical harmonics as~\cite{CrossGordonJCP66}:
\begin {multline}
\label{potential}
	V_\text{dd}(r)=- 8 \pi^{3/2} \left( \frac{2}{15} \right)^{1/2}
\frac{1}{r^3} \\
\times	\sum_{\nu\lambda} (-1)^{\nu+\lambda} C(1,1,2;\nu, \lambda, \nu+\lambda) \\
\times	Y_{1\nu}\left(\theta_1,\phi_1\right) Y_{1\lambda}\left(\theta_2,\phi_2\right)
	Y_{2,-\nu-\lambda}\left(\theta,\phi\right),
\end {multline}
with $C(j_1, j_2, j ; m_1, m_2, m)$ the Clebsch-Gordan coefficients~\cite{VarshalovichAngMom, ZareAngMom}. Here $(\theta_1,\phi_1)$ and $(\theta_2,\phi_2)$ are the rotational coordinates
of molecules 1 and 2, and $(\theta,\phi)$ are the polar coordinates of their relative position vector, $\hat{\mathbf r}$. The dipole-dipole interaction only mixes rotational states with $J'=J\pm1$ which have opposite parity.
\\[3pt]
\subsection{Induced dipole-induced dipole retarded interaction}

Far-off-resonant laser light induces oscillating dipole moments on each of the two molecules, and the retarded interaction between these instantaneous dipoles leads to an additional term in the intermolecular potential~\cite{CraigThiruBook, ThiruMolPhys80, SalamPRA07}:
\begin{equation}
	\label{VlaserInd}
	V_{\alpha \alpha} (k, \mathbf{r}) = \frac{I}{4 \pi \varepsilon_0^2 c} e_i^\ast \alpha^{\text{lab},1}_{ij}(k) V_{jl}(k,\mathbf{r}) \alpha^{\text{lab},2}_{l n}(k) e_n \cos(\mathbf{k r}),
\end{equation}
with $ V_{jl}$ the retarded induced dipole-induced dipole interaction tensor,
\begin{multline}
	\label{Vjk}
	V_{jl}(k,\mathbf{r}) = \frac{1}{ r^3} \bigl [ (\delta_{jl} - 3 \hat{\mathbf{r}}_j  \hat{\mathbf{r}}_l ) (\cos kr + kr \sin kr)  \\ -  (\delta_{jl} - \hat{\mathbf{r}}_j  \hat{\mathbf{r}}_l ) k^2 r^2 \cos kr \bigr ].
\end{multline}

For the laser light linearly polarized along the $Z$ axis, the retarded interaction~(\ref{VlaserInd}) can be rewritten in dimensionless form,
\begin{equation}
	\label{VlaserIndred}
	V_{\alpha \alpha} (k, \mathbf{r}) = \frac{\Delta \eta (k)}{\xi(k)}  \tilde{\alpha}^{\text{lab},1}_{Z j}(k) V_{jl}(k,\mathbf{r}) \tilde{\alpha}^{\text{lab},2}_{l Z}(k) \cos(\mathbf{k r}),
\end{equation}
with the energy measured in units of $B$, distance in units of $r_0$, $k$ in units of $r_0^{-1}$; $\tilde{\alpha}_{ij} = \alpha_{ij}/\Delta \alpha$ with $\Delta \alpha = \alpha_\parallel - \alpha_\perp$ is the reduced polarizability tensor. The dimensionless parameter,
\begin{equation}
	\label{Xiparam}
	\xi(k) = \frac{d^2}{2 \Delta \alpha(k) B},
\end{equation}
characterizes the relative strength of the permanent-dipole and induced-dipole interactions and is found to be on the order of $10^2-10^3$ for polar alkali dimers. 

We note that in eq.~(\ref{VlaserInd}) we neglected a ``static'' term due to the coupling of the dipole moment of one molecule with the hyperpolarizability of another via the optical field~\cite{BradshawPRA05, SalamPRA07}. This interaction is independent of $k$ and only becomes comparable to the dipole-dipole potential (\ref{DDpot}) at much larger intensities ($I \gtrsim 10^{14}$ W/cm$^2$) than considered here, at which ionization of alkali molecules is inevitable.

Equation~(\ref{VlaserIndred}) can be written in terms of spherical harmonics as:
\begin{widetext}
\begin{multline}
	\label{VlaserIndSpher}
	V_{\alpha \alpha} (k, \mathbf{r}) = 4 \pi \frac{\Delta \eta (k)}{\xi(k)} \frac{ \cos(\mathbf{k r})}{r^3}  \sum_{j,k} \underset{\mu_1, \mu_2} {\sum_{\lambda_1, \lambda_2}} F^{Zj}_{\lambda_1 \mu_1} F^{kZ}_{\lambda_2 \mu_2} \left[ (2\lambda_1+1)(2\lambda_2+1)  \right]^{-1/2} A^{\lambda_1}_0 A^{\lambda_2}_0 Y_{\lambda_1 \mu_1} (\theta_1, \phi_1) Y_{\lambda_2 \mu_2} (\theta_2, \phi_2) \\
	\times \left[ - \frac{2}{3} a(kr) \delta_{jk} \delta_{\mu_2, -\mu_1} + [a(kr)-3b(kr)] \sqrt{\frac{8\pi}{15}} F^{jk}_{2, -\mu_1-\mu_2} Y_{2, -\mu_1- \mu_2} (\theta, \phi) \right]
	\end{multline}
\end{widetext}
where the summation includes $j,k=1,2,3$; $\lambda_1, \lambda_2 = 0,2$; and $\mu_1, \mu_2 = -1, 0, 1$, and the coefficients are given by:

\begin{equation}
	\label{abCoef}
	a(kr) = k^2 r^2 \cos kr, \hspace{0.5cm} 	b(kr) = (\cos kr + kr \sin kr)
\end{equation}
The  coefficients $F^{jk}_{\lambda, \mu}$ are given in Appendix~\ref{sec:coefs}. 

The only nonzero components of $A^\lambda_\mu$ are given by:
\begin{equation}
	\label{Alm}
	A^0_0 = -\sqrt{3} \frac{\alpha}{\Delta \alpha}, \hspace{0.5cm} A^2_0 = \sqrt{\frac{2}{3}},
\end{equation}
with $\alpha=(\alpha_\parallel + 2\alpha_\perp)/3$ the average molecular polarizability. The optically-induced interaction~(\ref{VlaserIndSpher}) between molecules 1 and 2 mixes molecular states with $J'=J; J\pm2$ and $M'=M; M\pm1$ (the states of same parity) for each molecule.

\section{Combined effective potential}
\label{Sec:EffPots}

\begin{figure}[b]
\includegraphics[width=8cm]{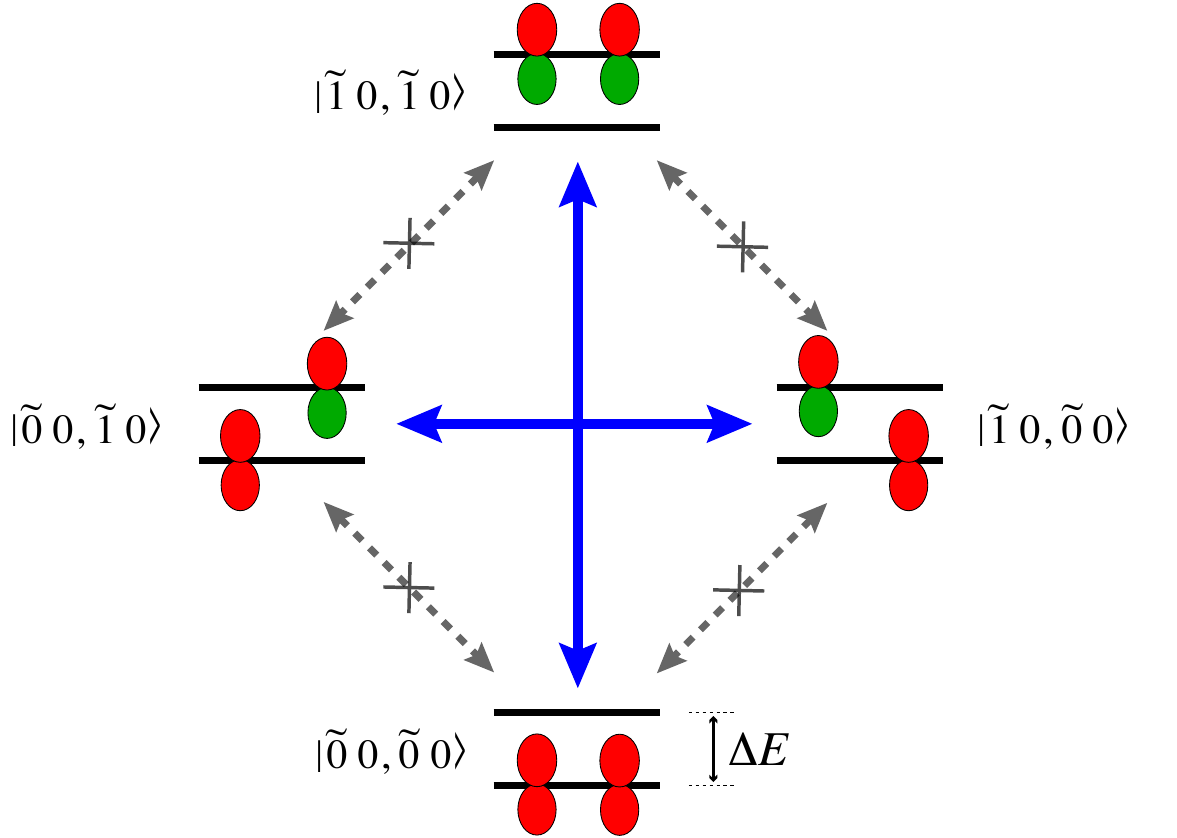}
\caption{\label{fig:levels} Schematic of the two-level configurations of interacting molecules. The pendular wavefunctions of laser-dressed molecules are shown schematically, with red and green colors indicating their sign. Blue arrows show the non-vanishing dipole-dipole interaction. See text.}
\end{figure}

Within the Born-Oppenheimer  approximation, the effective interaction potentials $V_\text{eff}(\mathbf{r})$ are obtained by diagonalizing the Hamiltonian for a fixed intermolecular separation $\mathbf{r} = (r, \theta, \phi)$,
\begin{equation}
	\label{Hamil}
	H = H_1 + H_2 + V_\text{dd} + V_{\alpha \alpha}, 
\end{equation}
where $H_1$ and $H_2$ are Hamiltonians (\ref{Halpha}) for molecules 1 and 2. In dimensionless units, the dipole-dipole potential~(\ref{DDpot}) is on the order of unity, while the strength of the optically-induced interaction~(\ref{VlaserIndred}) is given by $\Delta \eta/\xi$. Therefore, for a field strength parameter $\Delta \eta$ satisfying the inequality $1\ll \Delta \eta^{1/2} \ll \xi(k)$, both interaction terms are much smaller than the energy gap between neighboring tunneling doublets. Hence, in the basis of field-dressed states, $\vert \tilde{J}_1 M_1, \tilde{J}_2 M_2 \rangle$, the interaction between two ground-state molecules can be treated within the lowest tunneling doublet, $\vert \tilde{0}0, \tilde{0} 0 \rangle$--$\vert \tilde{1}0, \tilde{1} 0 \rangle$, as shown schematically in Fig.~\ref{fig:levels} (the $\vert \tilde{1}0, \tilde{0} 0 \rangle$ and $\vert \tilde{0}0, \tilde{1} 0 \rangle$ states do not interact with either the $\vert \tilde{0}0, \tilde{0} 0 \rangle$ or $\vert \tilde{1}0, \tilde{1} 0 \rangle$ states).  

Given that the optically-induced potential~(\ref{VlaserIndred})  and the dipole-dipole interaction (\ref{DDpot}) mix only states of same and opposite parity, respectively, the Hamiltonian matrix takes the form:
\begin{equation}
	\label{Hmatr}
	H  =  \left( \begin{array}{cc} U_{\alpha \alpha}^{\tilde{0}} & U_\text{dd} \\  U_\text{dd} & U_{\alpha \alpha}^{\tilde{1}} + 2 \Delta E  \end{array} \right),
\end{equation}
where $\Delta E (\Delta \eta)$ is the splitting between the tunneling-doublet levels $\vert \tilde{0}, 0 \rangle$ and $\vert \tilde{1}, 0 \rangle$, cf. Fig.~\ref{fig:AC_Stark},  and the matrix elements are given by:
\begin{equation}
	\label{Udd}
	U_\text{dd} (\mathbf{r}) = \frac{1- 3 \cos^2 \theta}{r^3} G(\Delta \eta),
\end{equation}
\begin{multline}
	\label{Uind}
	U_{\alpha \alpha}^{\tilde{J}}  (\mathbf{k}, \mathbf{r}) = \frac{\Delta \eta (k)}{\xi(k)}   \frac{\cos{(\mathbf {k r})}}{r^3} K_{\tilde{J}} (\Delta \eta, \alpha/\Delta \alpha) \\
	\times \left \{ -\sqrt{\frac{2}{3}} a (kr) 	+ \left [a(kr) - 3b(kr) \right ] \sqrt{\frac{8 \pi}{15}} Y_{20} (\theta, \phi) \right  \},
\end{multline}
with $\alpha = (\alpha_\parallel + 2 \alpha_\perp)/3$ the average molecular polarizability, and $a(kr)$ and $b(kr)$ given by eq.~(\ref{abCoef}).

The explicit expressions for $G(\Delta \eta)$ and $K_{\tilde{J}} (\Delta \eta, \alpha/\Delta \alpha)$ are given in Appendix~\ref{sec:matrel}. Their values are on the order of unity and in the strong-field limit, $\Delta \eta \to \infty$, take the analytic form:
\begin{equation}
	\label{Gstrong}
	G(\Delta \eta) = \left[ 1 - \Delta \eta^{-1/4} F \left (\tfrac{1}{2}\Delta \eta^{-1/4} \right) \right]^2,
\end{equation}
\begin{equation}
	\label{Kstrong}
	K_{\tilde{J}=0,1} (\Delta \eta) =  \sqrt{\frac{2}{3}}  \left[ \frac{\alpha_\parallel}{\Delta \alpha} - \Delta \eta^{-1/4} F \left (\Delta \eta^{-1/4} \right) \right]^2,
\end{equation}
where $F(x) = \exp (-x^2) \int_0^x \exp (y^2) dy$ is Dawson's integral~\cite{AbramowitzStegun}.

Therefore, the effective potentials are given by the eigenvalues of Hamiltonian (\ref{Hmatr}):
\begin{multline}
	\label{Veff}
	V_\text{eff} (\mathbf{k}, \mathbf{r}) = \Delta E + \frac{U_{\alpha \alpha}^{\tilde{0}} (\mathbf{k}, \mathbf{r}) + U_{\alpha \alpha}^{\tilde{1}} (\mathbf{k}, \mathbf{r})}{2} \\
	\pm \frac{\sqrt{4 U_\text{dd}^2 (\mathbf{r})  + \left[ U_{\alpha \alpha}^{\tilde{1}} (\mathbf{k}, \mathbf{r}) - U_{\alpha \alpha}^{\tilde{0}} (\mathbf{k}, \mathbf{r}) + 2 \Delta E \right]^2}}{2},
\end{multline}
with the minus sign corresponding to the ground state $\vert \tilde{0} 0, \tilde{0} 0 \rangle$ and the plus sign to the excited state $\vert \tilde{1} 0, \tilde{1} 0 \rangle$ effective potential.

\begin{table}
\centering
\caption{Parameters of the $^{40}$K$^{87}$Rb and $^{85}$Rb$^{133}$Cs molecules calculated from the data of Refs.~\cite{AldegundePRA08, KotochigovaDeMille10, DeiglmayrDulieuJCP08, AymarDulieuJCP05}. The intensity per unit strength parameter, $I/\Delta \eta$, is given in units of 10$^7$ W/cm$^2$. See text.}
\vspace{0.2cm}
\label{table:MolParam}
\begin{tabular}{| c | c | c | c | c | }
\hline 
& \multicolumn{2}{| c | }{KRb} & \multicolumn{2}{| c | }{RbCs}   \\
\hline
            & 1090 nm & static & 1090 nm & static \\
\hline
 $\xi$ & 96.5 & 434 & 572.0 & 3510\\
 $I/\Delta \eta$ & 1.4 & 6.5 & 0.43 & 2.7 \\
 $\alpha/\Delta \alpha$ & 0.62 & 1.38 & 0.58 & 1.52\\
  $\Delta \eta_c$ & 70 & 335 & 445 & 2812\\
 \hline
 $r_0$ [\AA] & \multicolumn{2}{ c  |}{36.1} & \multicolumn{2}{ c  |}{75.4} \\
 \hline
\end{tabular}
\end{table}

\begin{figure*}[t]
\includegraphics[width=15cm]{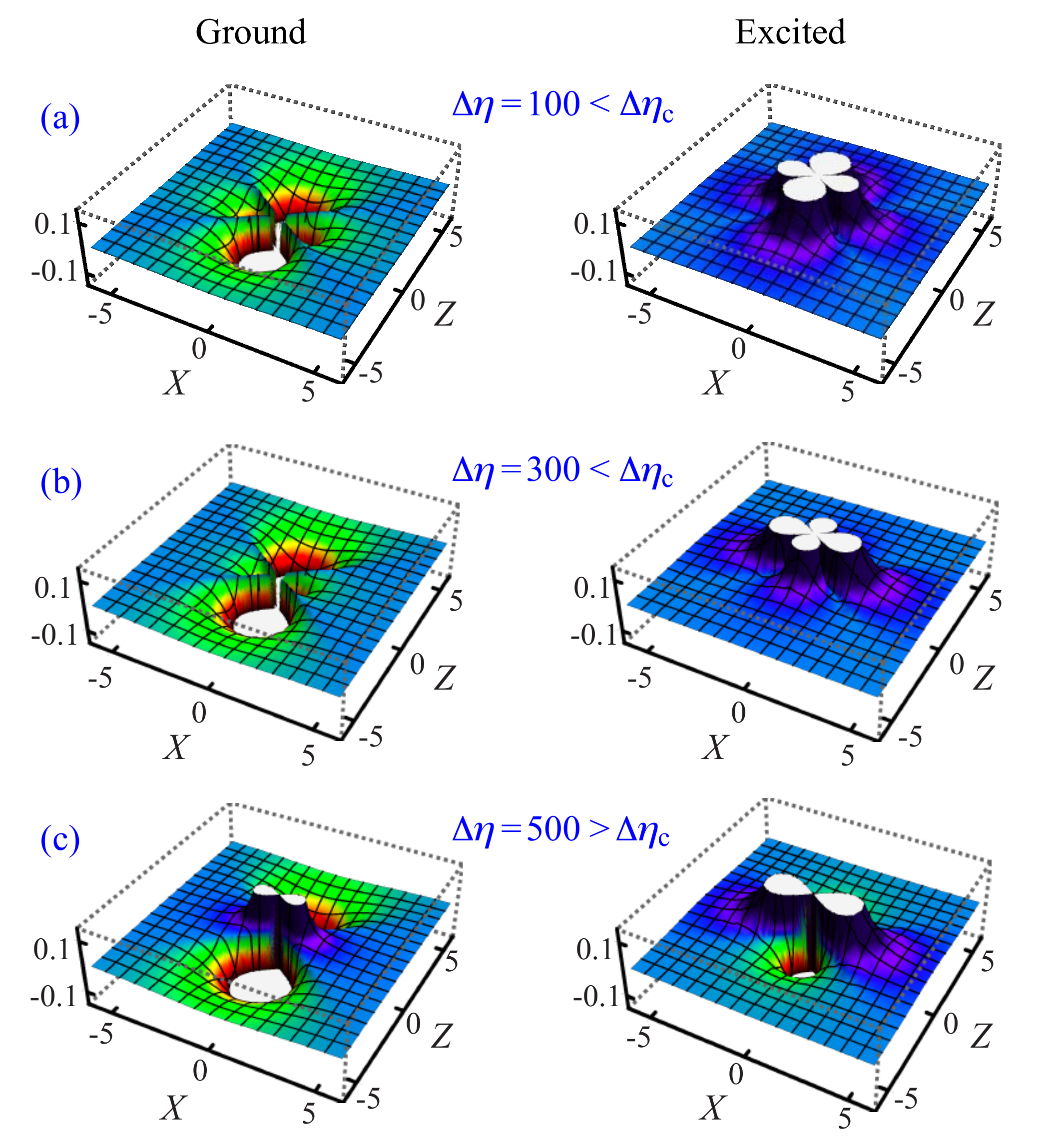}
\caption{\label{fig:RbCs_pots1090} Short-range behavior of optically-induced RbCs--RbCs potentials in the XZ plane ($\theta,\phi=0$), for different values of $\Delta \eta$ and a wavelength of 1090~nm. Left and right panels correspond, respectively, to the ground $\vert \tilde{0} 0, \tilde{0} 0 \rangle$ state, and excited $\vert \tilde{1} 0, \tilde{1} 0 \rangle$ state  potentials, as given by eq.~(\ref{Veff}). The dependence of the short-range potentials on $\phi$ is negligible. Potential energy is in units of $B$, with $V_\text{eff}(r \to \infty)$ chosen as zero; distances are in units of $r_0$. The laser beam propagates along the $Y$ axis, $\mathbf{k} \parallel \hat{\mathbf{Y}}$, with a polarization $\hat{\mathbf{e}} \parallel \hat{\mathbf{Z}}$. See text.}
\end{figure*}

\begin{figure*}[t]
\includegraphics[width=15cm]{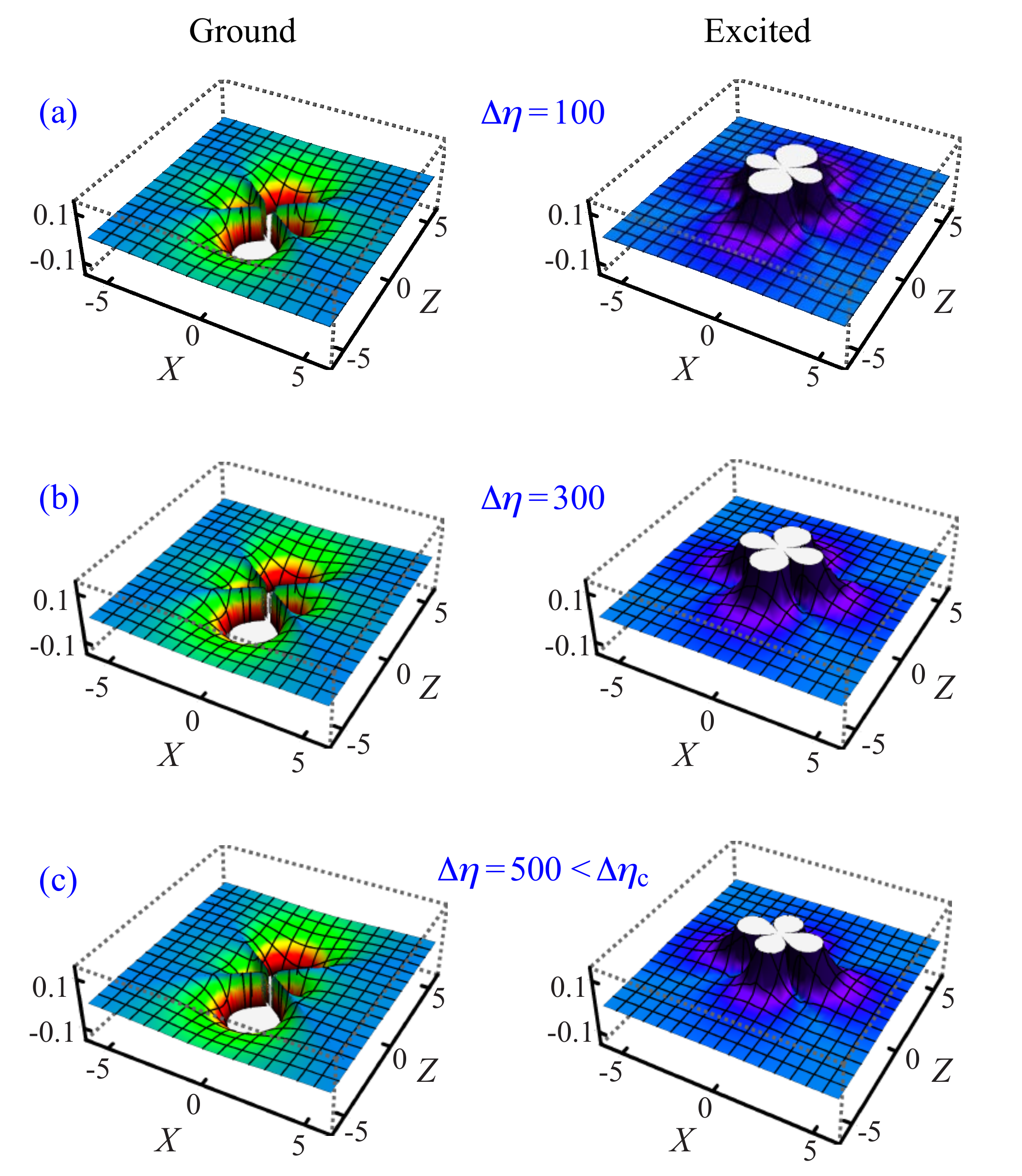}
\caption{\label{fig:RbCs_pots3000} Short-range behavior of optically-induced RbCs--RbCs potentials in the XZ plane ($\theta,\phi=0$), for different values of $\Delta \eta$ and a wavelength of  3000~nm. Left and right panels correspond, respectively, to the ground $\vert \tilde{0} 0, \tilde{0} 0 \rangle$ state, and excited $\vert \tilde{1} 0, \tilde{1} 0 \rangle$ state  potentials, as given by eq.~(\ref{Veff}). The dependence of the short-range potentials on $\phi$ is negligible. Potential energy is in units of $B$, with $V_\text{eff}(r \to \infty)$ chosen as zero; distances are in units of $r_0$. The laser beam propagates along the $Y$ axis, $\mathbf{k} \parallel \hat{\mathbf{Y}}$, with a polarization $\hat{\mathbf{e}} \parallel \hat{\mathbf{Z}}$. See text.}
\end{figure*}

\begin{figure*}[t]
\includegraphics[width=15cm]{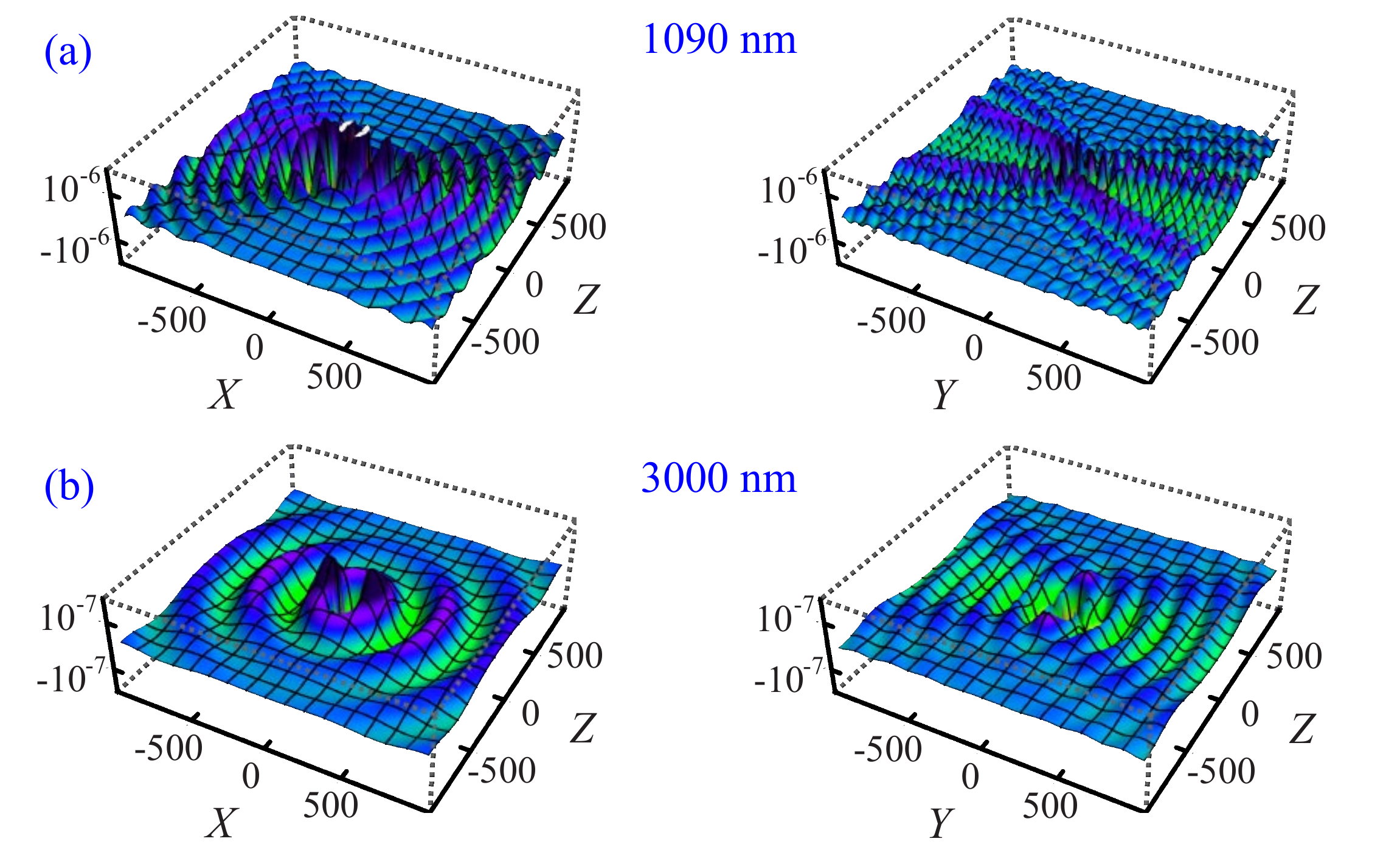}
\caption{\label{fig:RbCs_potsLong} Long-range behavior of the ground $\vert \tilde{0} 0, \tilde{0} 0 \rangle$ state RbCs--RbCs potentials in the $XZ~(\theta,\phi=0)$ and $YZ~(\theta,\phi=\pi/2)$ planes, for $\Delta \eta =100$ and wavelengths of 1090 nm (a) and 3000 nm (b). Long-range behavior of the excited $\vert \tilde{1} 0, \tilde{1} 0 \rangle$ state is similar; the magnitude of the oscillations scales with $\Delta \eta$, as given by eq.~(\ref{Vefflong}). Potential energy is in units of $B$, with $V_\text{eff}(r \to \infty)$ chosen as zero; distances are in units of $r_0$. Note different energy scales in panels (a) and (b). The laser beam propagates along the $Y$ axis, $\mathbf{k} \parallel \hat{\mathbf{Y}}$, with a polarization $\hat{\mathbf{e}} \parallel \hat{\mathbf{Z}}$.  See text.}
\end{figure*}

\begin{figure*}[t]
\includegraphics[width=15cm]{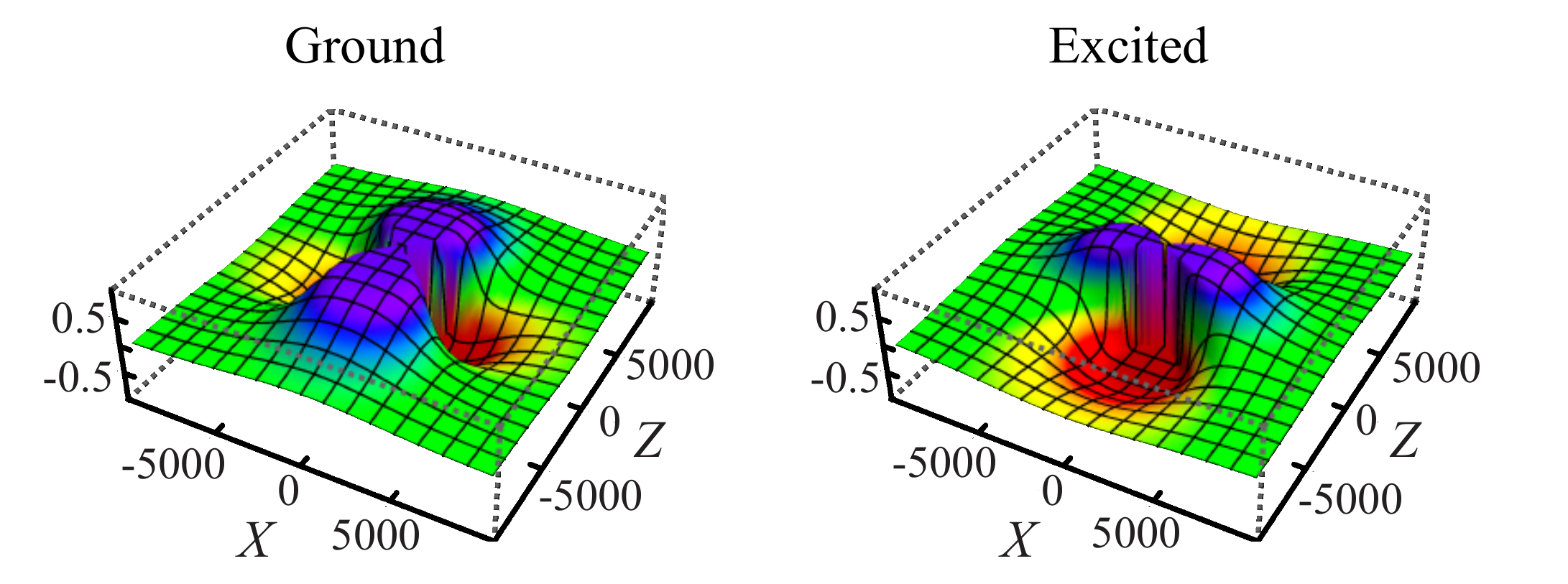}
\caption{\label{fig:RbCs_cos1090} Pair orientation cosines $\langle \cos \theta_{(1)} \cos \theta_{(2)} \rangle$ for an optically-dressed RbCs--RbCs pair in the XZ plane ($\theta,\phi=0$), for $\Delta \eta = 300$ and a wavelength of 1090~nm. Left and right panels correspond, respectively, to the ground $\vert \tilde{0} 0, \tilde{0} 0 \rangle$ state, and excited $\vert \tilde{1} 0, \tilde{1} 0 \rangle$ state  potentials, as given by eq.~(\ref{Veff}). The dependence of the short-range potentials on $\phi$ is negligible. The laser beam propagates along the $Y$ axis, $\mathbf{k} \parallel \hat{\mathbf{Y}}$, with the polarization $\hat{\mathbf{e}} \parallel \hat{\mathbf{Z}}$. See text.}
\end{figure*}

Table~\ref{table:MolParam} lists the values of the interaction parameters and the unit distance $r_0$ for RbCs as well as for KRb molecules, which are available in the laboratory, and serve herein as prototypical examples of ultracold polar molecules.
  
Figures~\ref{fig:RbCs_pots1090} and \ref{fig:RbCs_pots3000} show the short-range behavior of the effective potential of eq.~(\ref{Veff}), induced between a pair of $^{85}$Rb$^{133}$Cs  molecules by a laser field of wavelength $\lambda=1090$~nm and $\lambda=3000$~nm.

The behavior of the effective potential is dictated by the interplay between the static dipole-dipole and optically-induced dipole-dipole interactions, eqs.~(\ref{DDpot}) and (\ref{VlaserInd}). At small distances, $kr \ll 1$, the dominant contribution to the effective potential, eq.~(\ref{Veff}), becomes:
\begin{equation}
	\label{VeffShort}
	V_\text{eff} (kr \ll 1) \approx  \frac{\vert 1- 3\cos^2 \theta \vert}{r^3} \left[ \sqrt{\frac{3}{2}} \frac{\Delta \eta}{\xi} s(\theta) \pm G(\Delta \eta) \right],
\end{equation}
where $s(\theta) = \text{sgn} [1-3\cos^2\theta]$. Eq.~(\ref{VeffShort}) reveals the presence of a critical value, $\Delta \eta_c = \xi \sqrt{2/3} G(\Delta \eta)$, which determines the sign of the short-range potential  (the values of $\Delta \eta_c$ are listed in Table~\ref{table:MolParam}). For $\Delta \eta < \Delta \eta_c $, $V_\text{eff} (kr \ll 1)$ is governed by the second term in the square brackets, and so the potential is purely attractive in the ground state and purely repulsive in the excited state for any angle $\theta$, except for $\theta = \pm \arccos (\pm 3^{-1/2})$, where $V_\text{eff} (kr \ll 1)$ vanishes identically, cf. Fig~\ref{fig:RbCs_pots1090} (a), (b) and Fig.~\ref{fig:RbCs_pots3000}. This behavior is qualitatively different from the dipole-dipole interaction between two polar molecules oriented along the $Z$ axis,  $V_{dd}^{\uparrow \uparrow}=(1 - 3\cos^2 \theta)/r^3$, whose sign alternates in dependence on $\theta$. On the other hand, for  $\Delta \eta  > \Delta \eta_c $, the sign of $V_\text{eff}$ becomes angle-dependent, due to the interplay between the terms in the square brackets of eq.~(\ref{VeffShort}), resulting in a behavior similar to that of $V_{dd}^{\uparrow \uparrow}$: the potential is attractive at $\theta=0, \pi$ and repulsive at $\theta=\frac{\pi}{2}$, cf. Fig.~\ref{fig:RbCs_pots1090}  (c). Both below and above $\Delta \eta_c$, the inverse-power decay rate of $V_\text{eff}$ can be tuned by changing $\Delta \eta$. We note that whereas for RbCs and the non-resonant laser wavelength of 3000~nm  $\Delta \eta  > \Delta \eta_c $  corresponds to exceedingly high laser intensities, the requisite field strengths are substantially lower for KRb.


At large intermolecular separations, $kr \gg 1$, the optically-induced potential~(\ref{Uind}) becomes proportional to $1/r$ and thus dominates over the dipole-dipole interaction~(\ref{Udd}); its asymptotic behavior, given by,

\begin{equation}
	\label{Vefflong}
	V_\text{eff} (kr \gg 1) \approx - k^2 \frac{\Delta \eta}{\xi}  \sqrt{\frac{3}{2}} \frac{\cos ({\mathbf{k r}}) \cos (kr)}{r} \sin^2 \theta,
\end{equation}
becomes oscillatory, as shown in Fig.~\ref{fig:RbCs_potsLong} for the case of $\Delta \eta = 100$ and various laser wavelengths. The amplitude of the oscillations tapers off with increasing $r$.

Although at large intermolecular separations the optically-induced potential is on the order of $10^{-6}B$ ($\sim 25$ nK for RbCs), the $1/r$-like interaction, eq.~(\ref{Vefflong}), is of longer range than both the van der Waals and $V_{dd}^{\uparrow \uparrow}$ potentials, which may lend such optically-dressed molecules intriguing scattering properties.

In general, the effective potential of eq.~(\ref{Veff}) depends on the azimuthal angle $\phi$, as given by the $\cos (\mathbf {k r})$ term of eqs.~(\ref{Uind}) and (\ref{Vefflong}). While at short range the $\phi$-dependence is negligible, the long-range behavior of the effective potential is strongly anisotropic in $\phi$, cf. Fig.~\ref{fig:RbCs_potsLong}. As implied by eq.~(\ref{Vefflong}), the magnitude and phase of the long-range oscillations scale with $\Delta \eta$ and $k$ respectively, and are similar for the ground and excited states.

As displayed in Fig. \ref{fig:RbCs_potsLong}, the optically-induced potential exhibits concentric minima which, if deep enough, will support long-range bound states whose properties would be determined solely by the optical field, molecular dipole moments and polarizabilities -- and independent of the details of the intermolecular potential, in a manner reminiscent of the electrostatically-induced bound states predicted by Avdeenkov and Bohn~\cite{AvdeenkovBohnPRL03}. Analyzing the properties of these long-range states is a challenging theoretical and computational problem.

\section{Pair orientation cosines}

The dipole-dipole interaction, eq. (\ref{potential}), between a pair of molecules couples their tunneling doublet levels created by the laser field, thereby generating a parallel or anti-parallel orientation of the molecular dipoles.

In the laboratory frame, the  individual dipoles are not oriented, as implied by the vanishing values of their orientation cosines, $\langle \cos \theta_{(1,2)} \rangle=0$. However, the pair orientation cosine,  $\langle \cos \theta_{(1)}  \cos \theta_{(2)} \rangle \equiv \langle \psi_{g,e} \vert  \cos \theta_{(1)}  \cos \theta_{(2)} \vert \psi_{g,e}  \rangle$ does not vanish in general. Here $\vert \psi_{g,e}  \rangle = a_{g,e} \vert \tilde{0} 0, \tilde{0} 0 \rangle + b_{g,e} \vert \tilde{1} 0, \tilde{1} 0 \rangle$ are the ground and excited state eigenfunctions of Hamiltonian (\ref{Hmatr}).

Figure~\ref{fig:RbCs_cos1090} shows the pair orientation cosines for the optically-dressed RbCs--RbCs pair. One can see that as the molecules interact, they instantaneously orient each other even at very large distances on the order of 10 $\mu$m, due to a near-degeneracy of the tunneling doublet levels which is independent of the pair's separation. This corresponds to a long-distance orientational entanglement of the two polar molecules.

Another notable feature of the pair orientation displayed in Fig.~\ref{fig:RbCs_cos1090} is the strong dependence of the pair orientation cosines on the angle $\theta$ between the intermolecular axis $\hat{\mathbf r}$ and the polarization vector $\hat{\mathbf{e}}\parallel \mathbf Z$. Whereas for the intermolecular axis parallel to the laser polarization vector ($\theta =0$) the dipoles tend to orient each other in the same/opposite way for the ground/excited state, the sense of the mutual orientation in the ground and excited state reverses for the intermolecular axis perpendicular to the polarization vector ($\theta=\frac{\pi}{2}$). Thus the molecules of the pair ``hold'' each other ``up'' or ``down'' in either state depending on the polarization of the optical field.

\section{Conclusions}
\label{Sec:Conclusions}

In summary, we undertook a study of the interaction between a pair of polar molecules in the presence of an intense far-off-resonant optical field, and provided simple analytic expressions for the resulting potential energy surfaces. We found that the optically-induced potential is highly controllable and qualitatively different from the dipole-dipole interaction taking place between oriented polar molecules~\cite{BaranovPhysRep08}. The ability to engineer such potentials in the laboratory may open access to novel quantum phases in laser-dressed ultracold polar gases and to provide new methods to control molecular collisions in the ultracold regime. With a proper choice of tuning parameters, the dependence of the orientational entanglement of the pair is found to exhibit a weak, $1/r$ fall off with intermolecular distance $r$, lending the pair a particularly long-range entanglement: e.g., for RbCs, the pair orientation cosine vanishes only at intermolecular separations on the order of 10 $\mu$m.

\section{Acknowledgements}
We thank Eugene Demler, Mikhail Lukin, Boris Sartakov, and Timur Tscherbul for  insightful discussions; Svetlana Kotochigova for providing dynamic polarizabilities; and Gerard Meijer for encouragement and support.

\clearpage

\appendix

\begin{widetext}

\section{Coefficients of the optically-induced potential}
\label{sec:coefs}

The $F^{jk}_{\lambda, \mu}$ coefficients are given by:
\begin{equation}
	\label{Fjkcoefs}
	F^{jk}_{\lambda, \mu} = \sum_{\gamma, \delta} C(1 1 \lambda; \gamma \delta \mu) U^*_{\gamma j} U^*_{\delta k},
\end{equation}
where $C(j_1 j_2 j; m_1 m_2 m)$ are Clebsch-Gordan coefficients~\cite{VarshalovichAngMom, ZareAngMom}, and $U_{\gamma j}$ is given by:

\begin{equation}
	\label{Umatr}
U_{\gamma j} =  \left( \begin{array}{ccc} -\frac{1}{\sqrt{2}} & -\frac{i}{\sqrt{2}} & 0 \\  0 & 0 & 1 \\ \frac{1}{\sqrt{2}} & -\frac{i}{\sqrt{2}} & 0 \end{array} \right) 
\end{equation}

\section{Matrix elements of the dipole-dipole and optically-induced interactions}
\label{sec:matrel}

The coefficients of eqs.~(\ref{Udd}) and~(\ref{Uind}) are given by:
\begin{multline}
	\label{KJ}
	K_{\tilde{J}} = \underset{ J_1' J_2'}{\sum_{J_1 J_2}} \left[ \frac{(2J_1+1)(2J_2 +1)}{(2J_1'+1)(2J_2' +1)} \right]^{1/2} c_{J_1 0}^{\tilde{J}, 0} (\Delta \eta )  c_{J_2 0}^{\tilde{J}, 0} (\Delta \eta )  c_{J_1' 0}^{\tilde{J}, 0} (\Delta \eta )  c_{J_2' 0}^{\tilde{J}, 0} (\Delta \eta ) \\
	\times \sum_{\lambda_1 \lambda_2} N(\lambda_1, \lambda_2) A_0^{\lambda_1} A_0^{\lambda_2} \left[ C(J_1 \lambda_1 J_1'; 0 0 0) C(J_2 \lambda_2 J_2'; 0 0 0) \right]^2,
\end{multline}
with
\begin{equation}
	\label{Nlambda}
	N(\lambda_1, \lambda_2) = (\lambda_1^2 + \lambda_1 -2) (\lambda_2^2 + \lambda_2 -2)  \sqrt{\frac{2 (2\lambda_1+1) (2\lambda_2+1) }{3  (2-\lambda_1)! (3+\lambda_1)! (2-\lambda_2)! (3+\lambda_2)!}}
\end{equation}

\begin{equation}
	\label{Gfull}
 	G(\Delta \eta) =  \underset{ J_1' J_2'}{\sum_{J_1 J_2}} A(J_1, J_1') A(J_2, J_2') c_{J_1 0}^{\tilde{0}, 0} (\Delta \eta )  c_{J_2 0}^{\tilde{0}, 0} (\Delta \eta )  c_{J_1' 0}^{\tilde{1}, 0} (\Delta \eta )  c_{J_2' 0}^{\tilde{1}, 0} (\Delta \eta )  C(J_1 1 J_1'; 0 0 0) C(J_2 1 J_2'; 0 0 0)
\end{equation}
and with
 \begin{align}
	\label{Acoef}
	A(J, J') =  \frac{J+1}{\sqrt{(J+1)(2J+3)}};&  \hspace{0.3cm} J'=J+1 \\
		        - \frac{J}{\sqrt{J (2J-1)}};&  \hspace{0.3cm} J'=J-1 \\
		        0;   \hspace{1cm}  & \hspace{0.3cm}  \text{otherwise}
\end{align}

\end{widetext}

\newpage
\bibliography{References_library}
\end{document}